# C-BLUE 3 PC : a photon counting multimegapixel visible CMOS camera

Jean-Luc Gach*, Isaure De Kernier, Philippe Feautrier

First Light Imaging S.A.S., Europarc Ste Victoire – Bât. 5, Route de Valbrillant –13590 Meyreuil France

## ABSTRACT

The photon counting imaging paradigm in the visible and the infrared comes from the very small energy carried by a single photon at these wavelengths. Usually to detect photons the photoelectric effect is used. It converts a photon to a single electron making it very difficult to detect because of the readout noise of the electronics. To overcome this there are two strategies, either to amplify the signal to make it larger than the readout noise (used in the so called gain or amplified detectors), or to lower the readout noise in a standard image sensor. For a long time, only amplified detectors were able to do some photon counting. Since the first photon counting systems in the visible, developed by Boksenberg and his collaborators in 1972, many groups around the world improved photon counting techniques. In the 2000's in the visible, EMCCDs (electron multiplying charge coupled devices) allowed to replace the classical image intensifier photon counting systems by solid state devices and improved a lot the QE. But EMCCDs suffer from several issues, and the most important of them is the excess noise factor which prevents to know the exact incoming number of photons in the case of multiple photons per pixel. In the infrared there was no equivalent to EMCCDs up to the development of e-APD sensors and cameras made with HgCdTe material (electron initiated avalanche photo diode). With an excess noise factor near 1 at low temperatures, photon counting is possible with these devices but only in the infrared. We will show that having excess noise factor prevents from being able to do multiple photon counting (quanta imaging) and the only solution is to lower the readout noise.

**Keywords:** Photon counting, e-APD, EMCCD, infrared, imaging, quanta imaging

*jeanluc.gach@first-light.fr; phone +33 442612920; www.first-light-imaging.com

## 1. INTRODUCTION

Photon counting devices appeared a long time ago, as soon as it was possible to amplify the photonic signal sufficiently to overcome the readout noise of the electronics. This introduction shows the evolution of photon counting techniques through the ages up to modern photon counting in the visible.

### 1.1 The photomultiplier tube, first photon counting device

The fluxes involved especially for sources of low emission in astronomy (but not only) are of the order of a few photons per hour, which would correspond to photocurrents of some zepto ($1.10^{-21}$) ampere, impossible to measure with the current state of technology. It was therefore necessary to amplify this extremely weak signal to be able to detect it. The first idea led to the photomultiplier tube, combining photoelectric effect and secondary electronic emission.

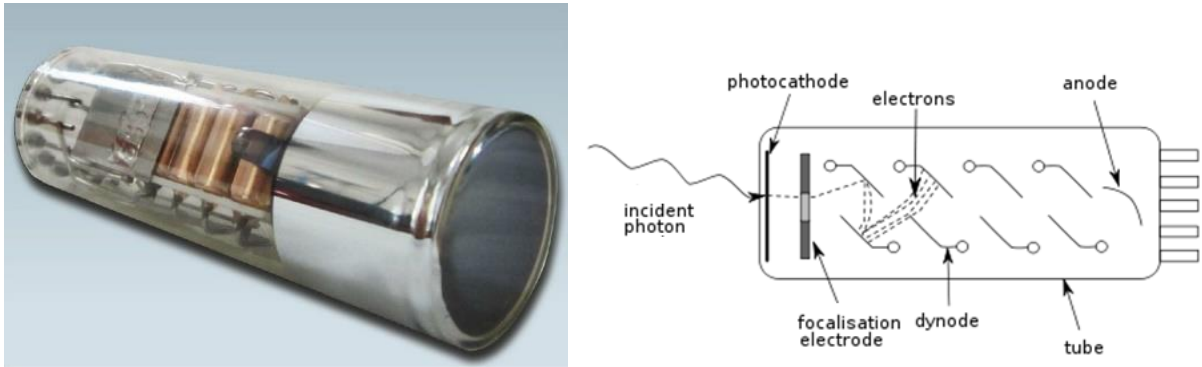

Figure 1 : Left, An image intensifier tube, right its principle using electron multiplication over dynodes in a vacuum tube.

The first tube was invented by NJ. Harley Iams and Bernard Salzberg of RCA in 1934 [1]. Figure 1 illustrates a photomultiplier tube embodiment and illustrates the principle of operation. An incident photon is transformed into a primary electron by a photocathode immersed in vacuum, this electron is accelerated and focused on a series of dynodes by an electrostatic field. During the impact of the electron on the dynode, one or more secondary electrons are emitted by ionization, these secondary electrons are in turn accelerated and focused by an electrostatic field on the next dynode. The anode collects the final jet of electrons and the measurement of the current coming from the anode makes it possible to measure the initial photonic signal amplified by the gain factor of the system which is essentially a function of the number of dynodes, their geometry and the field electric acceleration. This device is interesting since it shows the basic principles of photon counting that can be sliced in 3 independent processes, the photo-conversion done by the photocathode in this example, the amplification itself combining two different effects, electron acceleration in an electric field and secondary emission by impact ionization and finally current collection and detection on the anode of the tube.

### 1.2 Image intensified photon counting devices

Lallemand was the first to evoke the concept of bidimensional photon counting since he wrote in his 1936 historical publication [2] about the electronic camera (translated from French language): "This camera is a real photon counter, the photographic plate playing the role of totalizer of photon pulses at the same time as faithfully reproducing the point of arrival of each photon". The Lallemand camera was used up to the 80's. in astronomy (see figure 2 below).

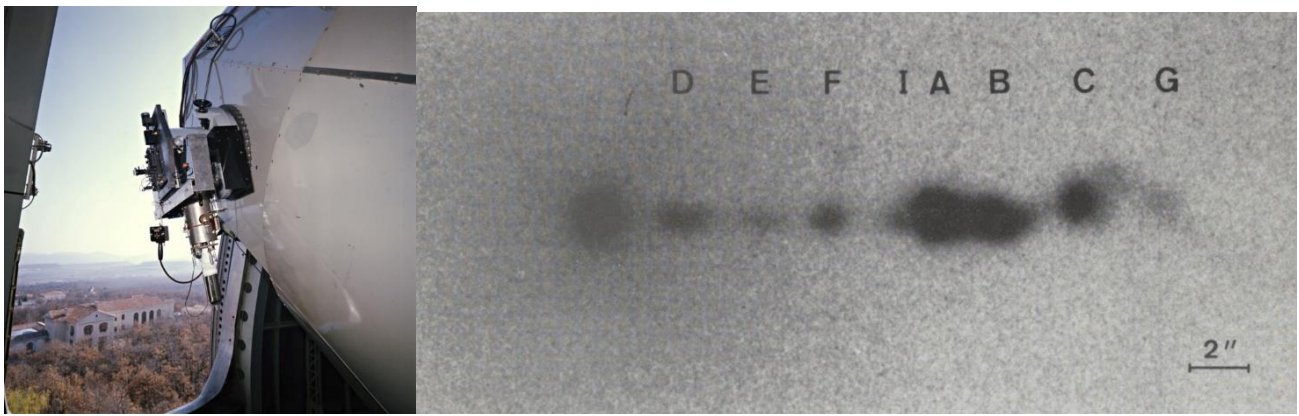

Figure 2: Left, the Lallemand camera used at the newton focus of the 193cm OHP telescope . (source photothèque CNRS-OHP). Right : High resolution image of M87 galaxy showing matter jets, obtained with a Lallemand camera at the 3.6m focus of the CFHT telescope in 1983 (Source gallica.bnf.fr / Bibliothèque Nationale de France)

However, it will be necessary to wait until 1971 and Gezari [3][4] to see appearing the first experiments of bidimensional photon counting by image intensifier tubes. In this case the image is no longer formed directly on the photo plate as in the Lallemand camera but on a phosphor screen. An optical fiber bundle coupling or lens system allows cascading intensifier tubes and thus obtain much higher amplification rates than those obtained by Lallemand and his electronic camera. Figure 3 shows a sectional view of a 2nd generation microchannel plate (MCP) intensifier tube. Figure 4 illustrates the operation of a microchannel plate. Each channel is a tube of 5 to 40 μm diameter covered with resistive material. A differential voltage is applied on each side of the MCP. An electron striking the wall will generate a number of secondary electrons accelerated by the electric field internal to the channel in the manner of dynodes. Each channel behaves like a micro photomultiplier tube.

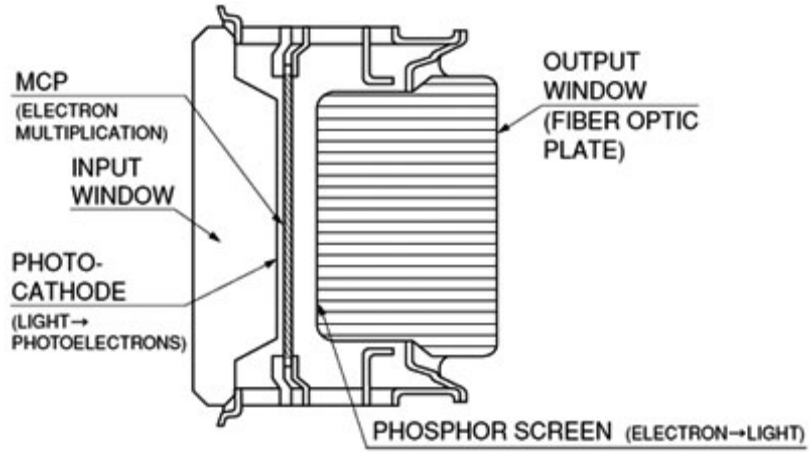

Figure 3: Structure of a microchannel plate (MCP) amplified image intensifier tube

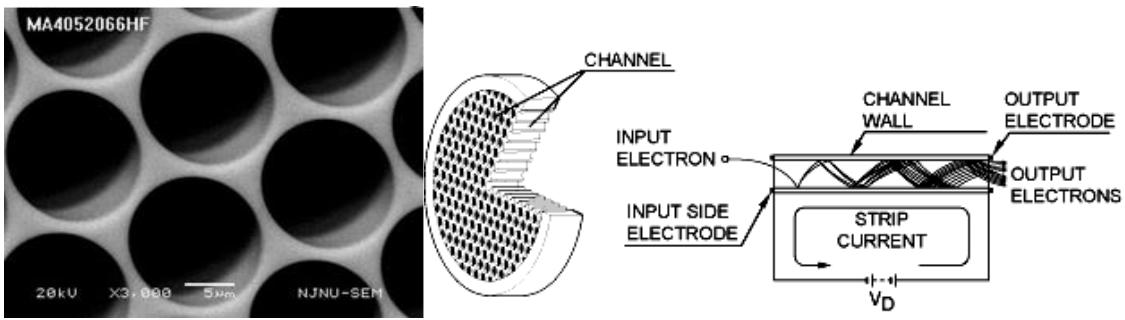

Figure 4: The MCP plate is a set of microtubes (SEM view left) coated with a resistive material acting as elementary micro photomultipliers and paving the plane to form a bidimensional image.

The Gezari device was described in more detail by Labeyrie et al. in 1974 [5], and mentioned the possibility of autocorrelations with a few photons per image. Figure 5 shows the detector used by Labeyrie and Gezzari. At the same time, Boksenberg understood from 1972 [6] the ability to detect single photons with this type of detector. He was also the first to couple an intensifier to an electronic reading system and not a photo plate (this was a plumbicon tube).

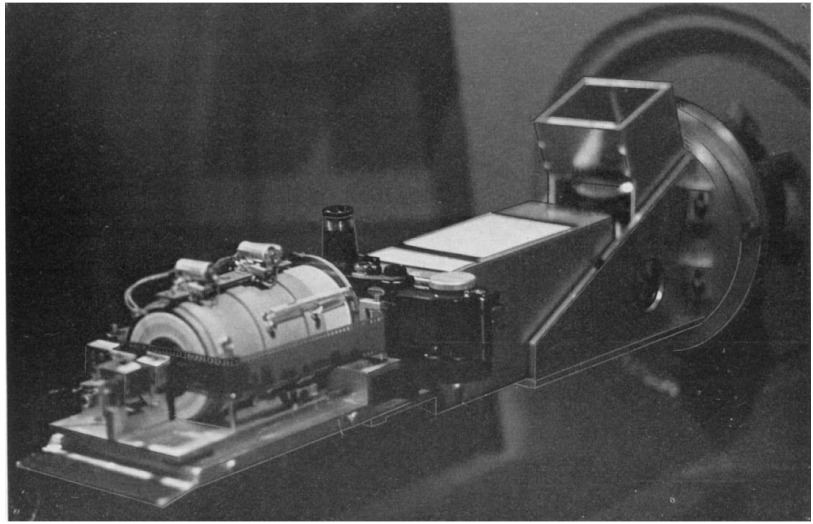

Figure 5: Detector employed by Labeyrie et al. to perform speckle interferometry experiments. The detector is a Kodak TriX film associated with a cascade of image intensifiers visible in the foreground.

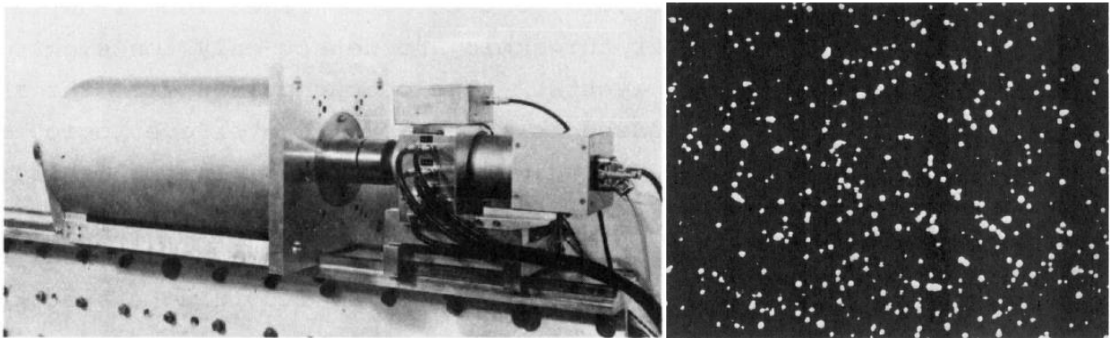

Figure 6 : Left: Boksenberg's photon counting camera. Right: a photon image produced by this camera.

In 1975 photon counting techniques were established and described in detail by Boksenberg [7]. The process differs slightly from standard imaging since the photon accumulation is not done on the imager itself but on a computer's memory. Each elementary image is sliced to a threshold level (usually $5\sigma_{RON}$), and the photons are detected by computing the center of gravity of each blob (see figure 7). Each image is a short time exposure and the accumulation is done by combining the positions detected in thousands of elementary images taken during the whole integration. The system has a big advantage since the slicing process is nonlinear it removes completely the noise in the image and therefore photon counting imagers have absolutely no readout noise. The price to pay is nonlinearity at high fluxes, since when several photons hit the detector at the same place on the same elementary image it is counted as one photon. This occurs also when two photos are spatially too close. This can be mitigated by avoiding the case using a fast readout and very short exposure times. This has been developed in Gach et al. [8].

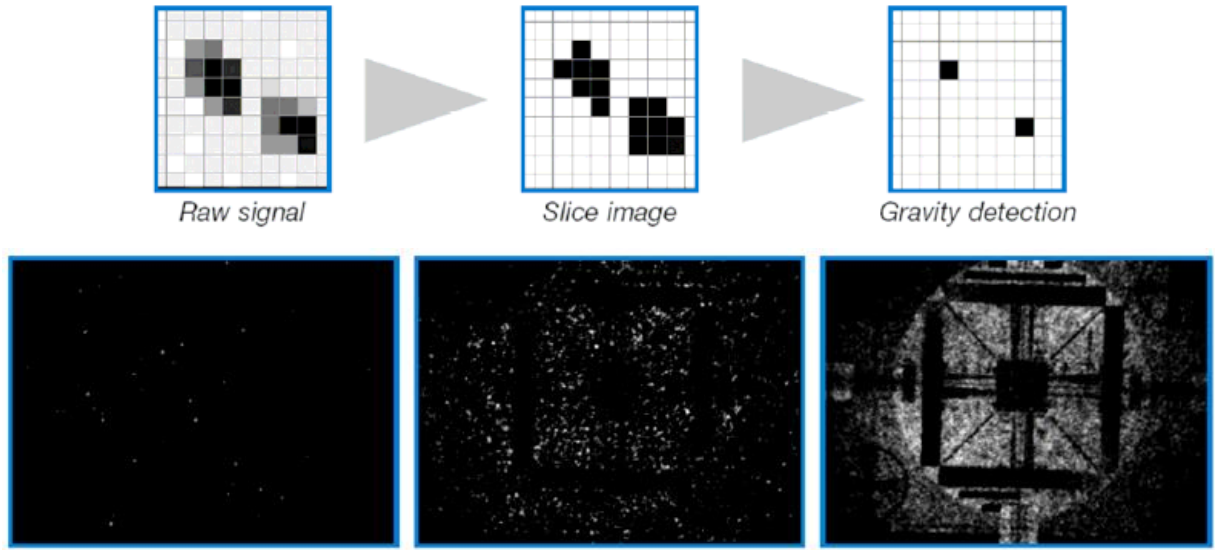

Figure 7 : Top : photon counting principle. the signal produced by an amplified photon in the image is sliced and a gravity computation is made to determine the initial photon impact localization. Bottom : an integration example, with an elementary 1/50$^{th}$ second integration image on the left, 10s accumulation reconstruction in the center and 5 min on the right (source Hamamatsu).

Subsequently the systems inspired by the one of Boksenberg were developed with refinements mainly concerning the acquisition which benefited from the advent of the computers. Thus Blazit [9][10] describes an autocorellation electronic system for speckle applications. In 1980 Boulesteix and Marcelin [11] describe a system with real-time acquisition, and showed that intensified photon counting systems (IPCS) were more efficient than very low flux charge coupled devices (CCDs). In 1995 & 1999 respectively EBCCDs (Williams et al. [12]) and EBAPS (Aebi et al. [13]) systems were introduced placing the detector directly in the vacuum tube and the avoiding the phosphor electron-to-photon conversion, in a sense coming back to the Lallemand camera, but with a electronic device instead of a photographic plate.

### 1.3 EMCCDs as photon counting devices

In 2002 the high quantum efficiency AsGa photon counting system was introduced (Gach et al. [8]). It was a real improvement of QE for such systems which were suffering from very poor quantum efficiency in the reddest part of the visible spectrum with classical bialkali or multialkali ($Na_2KSb$)Cs photocathodes. But it has been showed (for example by Boutboul [14]) that it is impossible to reach very high quantum efficiencies with such systems because on one side the photocathode must be thin to avoid phonon scattering and on the other side thick to maximize the absorption (Beer-Lambert law). Solid state systems were the only way to improve photon counting. In this sense the EMCCDs were a very promising path. The first avalanche multiplication experiments in CCDs (Charge Coupled Device) are reported very early in the literature by Madan et al. [15]. Various other authors have conducted subsequent experiments on the same theme, among which Gadjar & Burke in 1988 [16] or Hynecek in 1992 [17] and 1994 [18]. However, it will be necessary to wait for the work of Jerram et al. to see the first commercially available EMCCDs [19] from Marconi Applied Technologies or Hynecek at the same time [20] from Texas Instruments. EMCCDs coupled the advantages of solid state photoconversion with QE near 100%, an electron multiplication system and a detection circuitry on the same integrated device. As the multiplication process is done on a pixel per pixel basis, the gravity detection step of photon counting is no more necessary in this case. The first photon counting image with an EMCCD reported was in 2002 by Gach et al. [21] with a cryogenically cooled Marconi CCD65 (576x288 pixels) originally designed for TV night vision applications (figure 8).

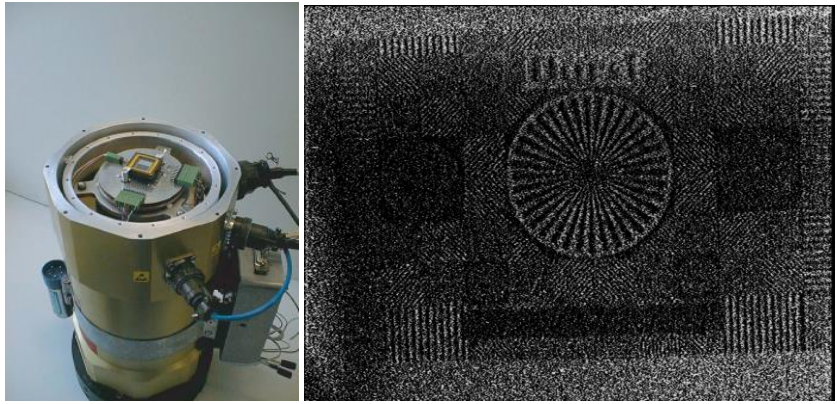

Figure 8 : The first EMCCD photon counting system reported in the literature in 2002 during the Scientific Detector Workshop [21], using a cryogenically cooled Marconi CCD65 (left) and a photon counting regime image at G=3000 (right).

As for image-tube-based photon counting, EMCCD photon counting does not permit to know the number of output electrons knowing the output amplified level. This is due to the stochastic nature of electron multiplication in the amplification register and therefore given an amplification level and input number of electrons, there is a probability to have an output level of electrons, and these probabilities overlap as plotted in figure 9.

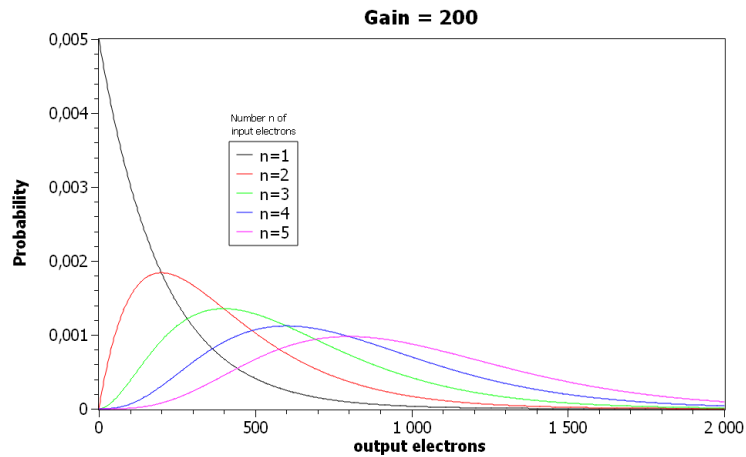

Figure 9: Output probability intensity of an EMCCD at a gain of 200 for various numbers of input electrons.

Basden and his collaborators showed in [22] that this probability can be expressed as:

$$P(x) = \frac{x^{n-1} e^{-\frac{x}{g}}}{g^n (n-1)!}$$

Where g is the mean gain and n the number of input electrons. This is the major limitation of photon counting EMCCDs, since one still needs to maintain a very high frame rate to overcome the nonlinearity at high fluxes due to photon coincidence or is obliged to use the sensor for very faint fluxes in the case of low framerates.

## 2. EXCESS NOISE FACTOR AND PHOTON COUNTING

**2.1 Theory of ionization**

McIntyre's theory [23] formalizes the stochastic effect of the process in various materials for avalanche triggered multiplication. His formalism uses the parameter k which is the ratio β/α of the ionization probabilities for holes and electrons per unit length, respectively. The theory allows to calculate the average gain <G> for a junction of length W as well as excess noise factor $F^2_{<G>}$ or ENF as a function of the average gain <G> that quantifies the stochastic effect of multiplication :

$$\langle G \rangle = \frac{(1-k)}{exp(-\alpha W(1-k)) - k}$$

$$F^2_{<G>} = k\langle G \rangle + (1-k)\left(2 - \frac{1}{\langle G \rangle}\right)$$

It will be noted that in the literature the excess noise factor is noted F or $F^2$ according to some authors.

Leveque et al. described a strong dissymmetry of k for Mercury Cadmium Telluride (MCT) in 1993 [25] and had already predicted a favorable situation for making avalanche photodiodes with a small ENF. Surprisingly, Beck et al. reported in 2001 an MCT MWIR avalanche photodiode structure whose excess noise does not follow McIntyre's theory [26][27]. The failure of McIntyre's theory can be explained in a highly dissymmetric material by the fact that an electron needs to acquire kinetic energy to be able to generate a new impact ionization, and this was not taken into account in McIntyre's theory where there is a non-null probability that a newly generated electron generates in turn immediately a new impact ionization. As early as 1974, Okuto and Crowell [28] hypothesized non-localized ionization coefficients, meaning that an electron that had recently passed through the conduction band could not in turn immediately generate a new impact ionization and had to acquire sufficient kinetic energy, which is obtained only after having traveled a given distance in the material, accelerated by the electric field. This model will then be identified in the literature as "hard dead space model". The impact of the introduction of this "dead space" is a self-organization of the multiplication process which then becomes localized in space at specific points and thus decreases the stochastic aspect of the process. Other authors will then develop this theory including Saleh in 1990 [29], a theory that will correctly predict the ENF of certain diodes (Hayat 2000 [30]). Other researchers worked in parallel to the refinement of this theory, this is the case of Marsland in 1990 [31]. The same author made a good summary of the state of these different variants in 2011 [32] and showed that all these works eventually lead to close results. Derelle and her collaborators have built a model involving not only the notion of "dead space" but also phonon and alloy scattering computed by Monte Carlo simulation in good agreement with the observations [33] made on diodes developed by Rothman and his collaborators [34]. Today the question does not seem completely closed, and teams continue to work on the subject with interesting results like those of Bertazzi et al. in 2010 [35] or Belotti et al. in 2011 [36].

**2.2 Proportional photon counting (quanta imaging)**

An excess noise factor close to 1 makes it possible finally to consider a count of photons in proportional mode and no longer by thresholding, in other words the possibility to go back to the initial number of photons knowing the number of electrons at the output of the amplified system. The advantage is obviously to overcome the problem of coincidence and maintain good linearity even at very high flux. The limit from which this differentiation is possible has been formalized in previous papers about e-APDs [37][38]. Figure 10 shows the output spectrum for an evenly distributed source of photons, and for different ENF, while figure 11 shows the spectrum on a real source, convolved with the Poisson distribution.

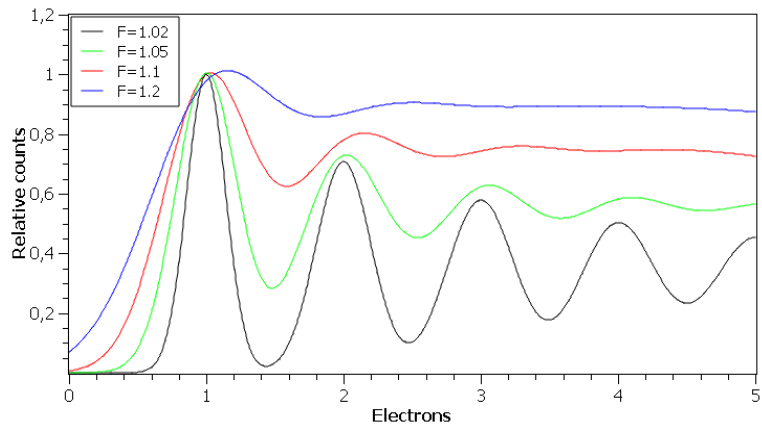

Figure 10: Standard output spectrum of an APD photodiode for a uniform distribution of the number of injected electrons and for various excess noise factors.

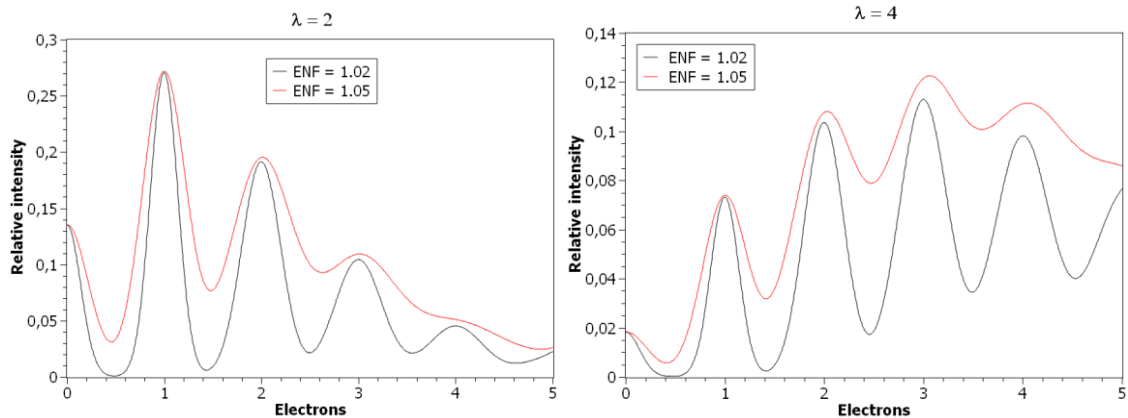

Figure 11: Normalized (input referred) spectrum for a 2-photon (left) and 4-photon (right) source by elementary integration for an excess noise factor of 1.02 and 1.05 and for an electronic noise σ = M / 5 (M being the multiplication gain)

The conclusion is that $F^2$ needs to be as low as 1.05 to start to see the 1 vs 2 photon intensity peaks, and 1.02 is preferable to be able to do proportional photon counting up to 5 simultaneous photons. One could speculate that the closest to 1 is the ENF, the most favorable is the situation for individual photons differentiation. Such ENF is only possible with detectors like e-APDs for the moment, therefore in the infrared. In the visible such detectors don't exist yet so it is necessary to lower the noise to be able to do quanta imaging instead of using amplified detectors.

## 3. PHOTON COUNTING WITH CMOS DEVICES

### 3.1 Basic principle

Fossum showed in 2013 [39] that it was necessary to have a readout noise lower than 0.3 electrons to be able to do proportional photon counting. Figure 12 shows the output histogram achieved with images sensors of various readout noise. The smaller the readout noise is, the better it is possible to discriminate individual events and avoid misclassifying the number of input photons. This curve is obtained by convolving a Poisson distribution with a gaussian with a width corresponding to the sensor's readout noise.

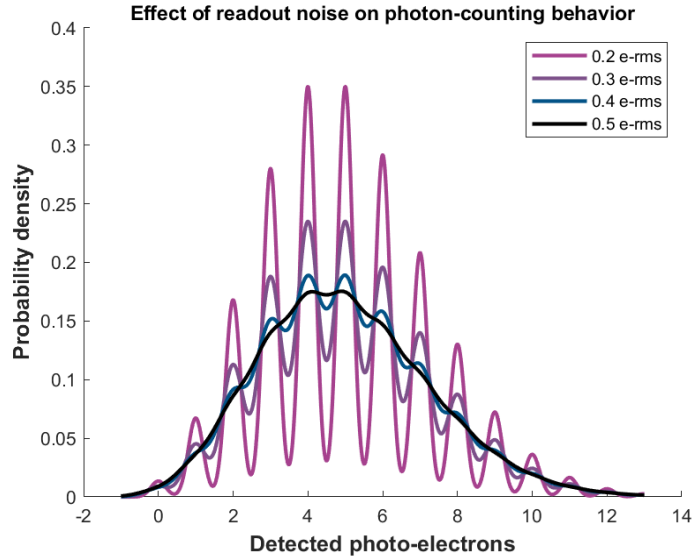

Figure 12: Simulation of the output histogram of an image sensor for various readout noises

The misclassification (bit error rate) can also be easily modeled and is a function of the readout noise. Figure 13 shows this bit error rate as a function of the readout noise.

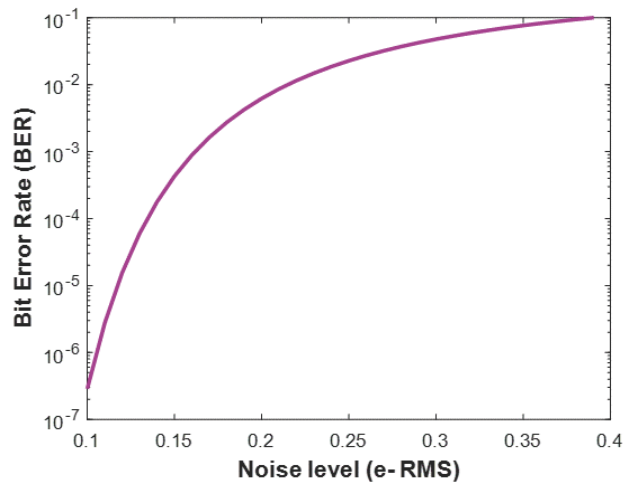

Figure 13 : Bit error rate as a function of the readout noise.

### 3.2 Achieving a low readout noise

There is several ways to obtain a low readout noise in a CMOS sensor. The first is to lower the intrinsic noise of the source follower of the pixel, this is a technological improvement. The other way is to lower the source follower node capacitance. By doing so, the charge to voltage conversion which is inversely proportional to the node capacitance is higher and the signal to noise ratio is increased accordingly. This can be done by reducing the pixel size and using a gateless reset transistor to avoid all the parasitic capacitances. Using small size transistors also helps. Masoodian and his collaborators reported the design of an 1.1μm pitch active pixel sensor whose gain is in the range of 350μV/e-, much higher than standard CMOS devices[40]. Later Ma et al. reported the successful manufacturing of a 16.7 Mpixels imager with a stacked structure for image readout [41]. Other strategies are also possible using more classical pixel structures exhibiting lower (but still high) conversion gains but using non destructive multiple readout to lower the overall readout noise, as reported by Stefanov et al. who claimed achieving 0.15 electron readout noise with non destructive readout [42].

## 3.3 The C-BLUE 3 PC camera

Table 1: C-BLUE 3 PC main features

| Sensor size | | 4096 x 4096 |
|---|---|---|
| Pixel pitch | | 1.1 µm |
| Shutter architecture | | Rolling shutter |
| Readout modes | | Photon-counting<br>Normal |
| Framerate (full frame) | | 10 to 30 Hz |
| Readout noise | Photon-counting | 0.22 e-MED<br>0.32 e-RMS |
| | Normal | 0.53 e-MED<br>0.58 e-RMS |
| Dark current (10°C) | | 0.002 e-/p/s |
| Full well | Photon-counting | 280 e- |
| | Normal | 2100 e- |

Based on the previous sensor, the camera attempt has the characteristics summarized in table 1. It is important to add that this camera will be able to do photon counting at full speed, near room temperature thanks to a very low dark current and will not exhibit the major defects of EMCCDs like clock induced charge (CIC) or transfer inefficiency as reported by Daigle et al.[43]. Figure 14 shows a sample image of the camera prototype in photon counting mode with its histogram showing perfectly discrete peaks in it at the digital number position of each photon count.

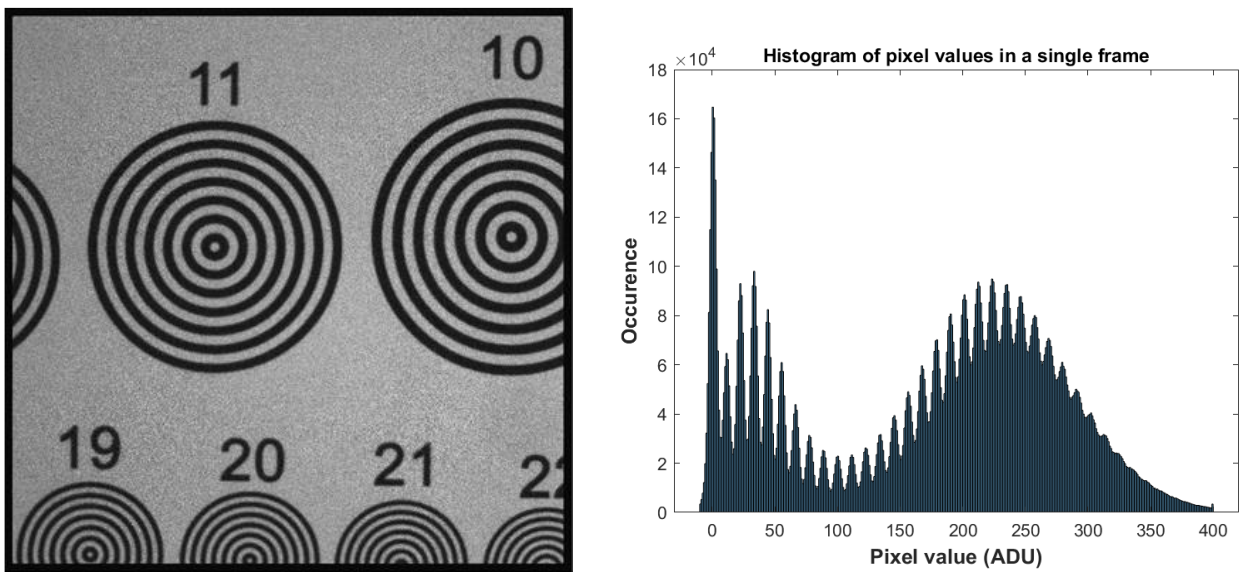

Figure 14: Left: image of the C-BLUE 3 PC prototype in photon counting mode. Right: histogram showing discrete peaks at the position of each photon count (note that the histogram bins blur the valleys)

### 3.4 An easy way of measuring system gain with quantum images

The usual way of measuring an imager's system gain is to use the well-known photon transfer curve. In the case of photon counting devices (quanta image sensors) it is simpler, because the peaks in the histogram are the track of an integer number of photons, and then the distance between two peaks maximum is the direct measurement of the gain. Of course, by using only two consecutive peaks, the error measurement can be important. Starkey and Fossum proposed to use the peaks position and make a linear fit to improve the measurement precision [44]. Another way of doing this measurement is to do the Fourier transform of the histogram to detect the peaks frequency. The maximum of the first peak in the single sided amplitude Fourier spectrum of the histogram gives directly the gain measurement (in electrons per analog digital unit or digital number) using several peaks simultaneously thanks to the measurement in the frequency domain. Figure 15 shows such a spectrum. It can be noted that this method does not need any specific experimental setup, it can be done on any image with a few photons per pixel.

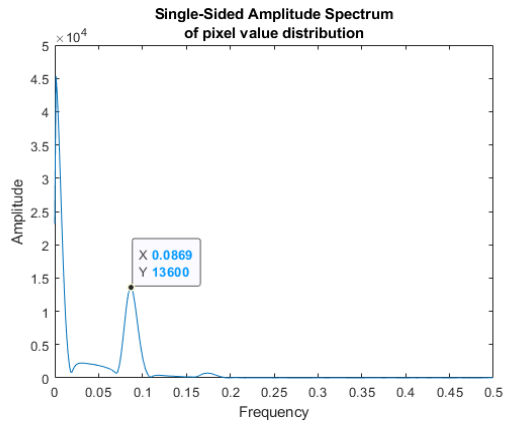

Figure 15 : single sided amplitude Fourier spectrum of the histogram in photon counting regime. The peak position gives directly the conversion gain value.

### 3.5 Readout noise

Figure 16 shows the readout noise distribution and readout noise map. With a mean readout noise of 0.32 electrons it is possible to achieve photon counting and quantum imaging. The noise image shows no excess noise in any part of the image whereas the histogram is asymmetric like seen in many CMOS imagers but contained to a very acceptable level.

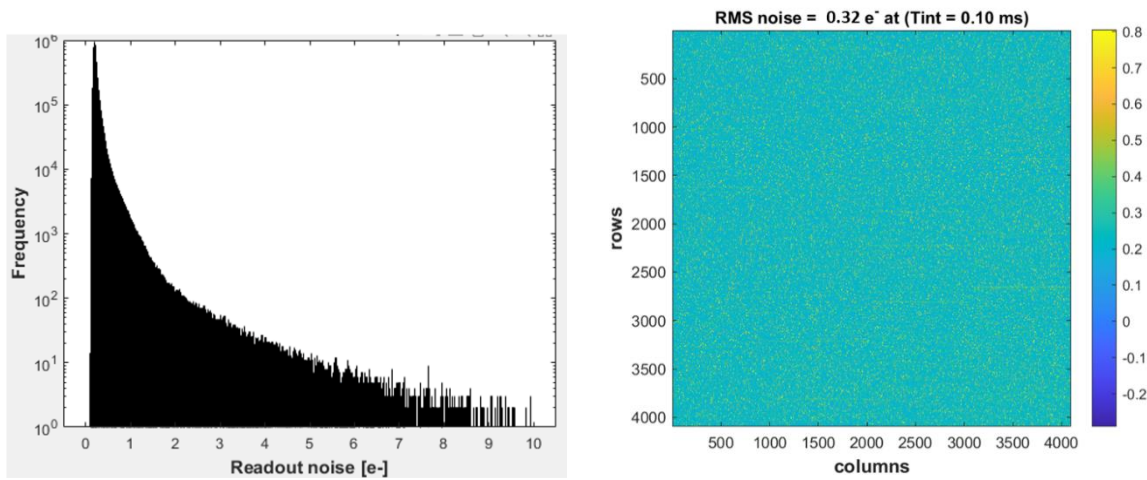

Figure 16: Noise image an noise histogram of the camera prototype, with 0.21 electron median noise, 0.25 electron mean and 0.32 electron RMS.

### 3.6 Quantum efficiency

The measured quantum efficiency is shown in figure 17. The sensor is front side illuminated and therefore needs some microlenses to reach acceptable QE. However with these microlenses it reaches a good level (peaks at more than 80%) and has a curve comparable to classical CMOS imagers. The acceptance angle of the imager has not been measured yet and will be in the near future.

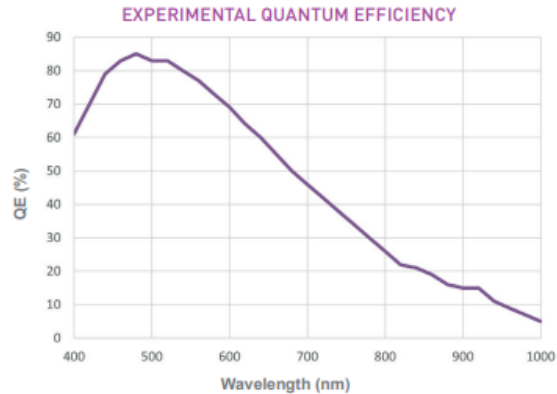

Figure 17: Measured quantum efficiency of the camera

## 4. ASTRONOMICAL OBSERVATIONS WITH QUANTUM IMAGING CAMERAS

### 4.1 Creating a quantum image

The raw image incoming from a camera such as C-BLUE 3 PC is nothing else than a very low noise CMOS image. The quantization process makes it become a quanta image. To do so, it is just necessary to do a multi-level thresholding of the image with threshold steps equal to the system gain determined for example in a method like described in 3.4 and adjusted so that the thresholding levels fall in a valley of the pixel histogram. It is necessary to do this independently for each pixel so to have a gain map for each pixel. The output of this process is a per pixel gain corrected quantified image containing only the photon numbers. It is interesting to notice that this process is nonlinear and completely suppresses the remaining readout noise in the image. The only added noise is the potential misclassification of one photon count to the next or the previous, the probability of this is the bit error rate as described in section 3.1. This introduces some extra noise but it has to be compared to the photon noise, therefore it plays a significant role only at extremely low light levels.

### 4.2 Temporal information

When images are free of readout noise, as described in the previous section, it is then possible to add them in a lossless way. By doing so the final image is strictly equivalent in terms of signal to noise ratio to the long exposure image that totalizes the exposure time of each individual image. This is not the case with standard imaging because you keep the readout noise that adds in the final image quadratically. Actually, in quanta images the readout noise also adds in a quadratic way but since the readout noise is null it ends to a zero term.

Because of this there is an interest to use the camera at its maximal framerate. Doing so one can adjust a posteriori the exposure time by adding the elementary images. This is interesting for example if someone wants to recover the instantaneous residual PSF after an AO system to do deconvolution, or to do image recentering on non-AO telescopes, thus increasing the spatial resolution.

### 4.3 Spatial information

In the same way blocks of NxM pixels can be added loslessly. So the pixel matching to the turbulence is no more an issue since it is possible to readjust a posteriori the spatial sampling to match the seeing variation or AO performance. In that case oversampling the PSF is no more a loss in performance since it is possible to have an optimal sampling by the a posteriori binning process. It would also be possible to build variable or multi resolution images, not using the same binning factor in regions where there is flux vs regions where the signal is weak and adapting the pixel sampling to the object. In the same way in spectrography where one dimension can be predominant, it is possible to build loslessly rectangular pixels at will.

It is counter-intuitive to what is done with standard CMOS devices where it is better to do binning at observation time and not a posteriori to minimize the noise. Here it is the contrary: if hardware binning is used, two phenomena will occur, the first is the quadratic sum of the readout noise of each pixel, thus increasing the binned pixel's noise to a level where photon counting is not possible. Secondly, the gain mismatch of each pixel will blur the quantum histogram because peaks corresponding to each quantum are not perfectly inline whereas the a posteriori binning is lossless because of the null readout noise in the image, in a similar manner as the temporal sum exposed in the previous section.

## 4.4 Dynamic range

In such a system, if the previous ways of observing are applied, in particular the preservation of the temporal information running the camera at its maximal framerate and summing the images at posteriori after thresholding to form the final image, the dynamic range (DR) of the image which is the maximal detectable signal without saturation divided by the minimal one (here one photon) is virtually infinite and proportional to the exposure time. It can be expressed as:

$$DR = \frac{signal_{max}}{signal_{min}} = \frac{FWC \times FR \times t}{1} = FWC \times FR \times t \qquad (1)$$

Where FWC is the full well capacity in photon counting mode, FR is the framerate in Hz and t is the total exposure time in seconds. For example with a sensor having a FWC=300 electrons in photon counting mode, running at 30Hz and using a 10 minute exposure time (600s) then the dynamic range is :

$$DR = 300 \times 30 \times 600 = 5\,400\,000\ (134 dB)$$

This number is to be compared to classical CMOS devices with high dynamic range capability that usually exhibit a dynamic range of 30 000 to 60 000 (90 to 96 dB). It is important however to maintain the single image signal to a level where saturation is not reached. This can be done by increasing the framerate of the sensor. It is important also to notice that at the contrary for very short exposure times the dynamic range of the sensor is very low. Inverting equation (1) gives the minimal exposure time required to obtain a given dynamic range. It easy then to see that to reach a standard dynamic range of 10 000 it is necessary to form an 1.1s exposure at 30Hz and with a FWC of 300.

## 5. CONCLUSION

We presented the first evaluation of a quantum image CMOS sensor and its possible application to astronomy. These sensors are very promising and show the path of low light or high dynamics imaging in astronomy superseding largely EMCCDs in that field. There is now several sources of quanta image sensors and we believe that many different other manufacturers will propose this functionality to their imagers since it is compatible with existing and largely used CMOS processes, benefiting of the rise in power of CMOS imagers as well.